\documentclass[aps,prc,showkeys,showpacs]{revtex4}
\usepackage{graphicx}
\usepackage{dcolumn} 
\usepackage{bm}
\usepackage{hyperref}
\usepackage{amsmath}
\usepackage{multirow}
\usepackage{color}
\begin{document}
\title{$2\nu\beta\beta$-decay to first $2^+$ state with partial isospin symmetry restoration from spherical QRPA calculations}
\author{Dong-Liang Fang$^{a,b,c}$ and Amand Faessler$^{d}$}
\address{{$^a$Institute of Modern Physics, Chinese Academy of Sciences, Lanzhou, 730000, China}}
\address{{$^b$Center for Theoretical Physics, Jilin University, Changchun, 130012, China}}
\address{{$^c$School of Nuclear Science and Technology, University of Chinese Academy of Sciences, Beijing, 100049, China}}
\address{$^d$Institute for theoretical physics, Tuebingen University, D-72076, Germany}
\begin{abstract}
With partially restored isospin symmetry, we calculate the nuclear matrix element for a special decay mode of $2\nu\beta\beta$ (two neutrino double beta decay) -- the decay to the first $2^+$ excited states. With the realistic CD-Bonn nuclear force, we analyze the dependence of the nuclear matrix elements on the iso-vector and iso-scalar parts of proton-neutron particle-particle interaction. The dependence on the different nuclear matrix element is observed and the results are explained. We also give the phase space factors with numerical electron wave functions and properly chosen excitation energies. Finally we give our results for the half-lives of this decay mode for different nuclei.
\end{abstract}
\keywords{ double beta decay, quasi-particle random phase approximation, nuclear matrix element}
\pacs{14.60.Lm,21.60.-n, 23.40.Bw}
\maketitle
\section{introduction}
Double beta decay (hereafter $\beta\beta$-decay) is a rare decay which happens under nuclear circumstance. Its possible mode called neutrinoless double beta decay (hereafter $0\nu\beta\beta$) provides insights of physics beyond Standard Model. Such a decay mode would give clear evidence of lepton number violation. However, the discovery itself is not enough to tell us the exact mechanism of  this violation. Therefore further investigations on the underlying physics behind are needed after the discovery of this rare process. Different methods have been proposed for probing these underlying new physics, such as comparisons among half-lives of various nuclei or between the decays to ground and excited states \cite{DLZ11}. Recent surveys show that the decays among different nuclei may be correlated \cite{BSK18} for selected mechanisms such as the light and heavy mass mechanisms. Meanwhile, the relative decay width to $2^+$ states may be a better way to distinguish between models with presence of right-handed weak currents \cite{DKT85,Tom86}. However, to describe such a process, one needs reliable and capable modern nuclear many-body methods. To test the reliability of these methods, we first apply our many-body calculations on double beta decay with neutrinos to the first $2^+$ excited states. These results will also help the search of such a mode from various collaborations \cite{Bar17}.

The double beta decay with neutrinos, named two-neutrino double beta decay (2$\nu\beta\beta$) transforms an even-even nucleus ${}^{A}X_N$ to a neighboring even-even nucleus ${}^{A}Y_{N-2}$ with the emission of two electrons and two anti-electron-neutrinos. Due to angular momentum conservation, the change of angular momentum for the decaying nucleus are sums of angular momenta of the four emitted leptons. For the $2\nu\beta\beta$, the electrons and neutrinos are dominated by s- partial waves, because they have relative long wavelength, compared to the nuclear radius, due to their small momenta. Therefore, for the leading order contribution to the decay, the summed angular momentum of the outgoing leptons from $\beta\beta$-decay, can have the value of $0$, $1$ and $2$ only. If we go through the nuclear chart, one finds that there are only a limited number of excited states of double beta decay daughter nuclei within the $Q_{\beta\beta}$-value windows. They have spin-parities $2^+$ and $0^+$ only. Thus, for many $\beta\beta$-decay candidates with large enough $Q_{\beta\beta}$ values, there exist possibilities of decays to the first $2^+$ states ($2^+_1$) of the daughter nuclei. All these decays are  suppressed by the large energy denominator \cite{DKT85} compared to $2\nu\beta\beta$ decaying to ground states. Therefore they have small branch ratios and a low detectability \cite{Bar17}. Precise predictions are then helpful for experimentalists. Despite these, measurements of these decays can help to improve nuclear theories and solve problems in nuclear structure calculations, such as the quenching of $g_A$. There have been already in the literatures \cite{HSB07,CS94,AS96,SM96,SSF97,SSF98,SCR17}, from Shell model and QRPA, investigations of this issue. But the deviations from model to model are still large, usually they differ several orders of magnitude. In this work, with the isospin restoration \cite{RF11,SRF13}, we systematically investigate this issue. By studying the NME dependence on the particle-particle (pp) interaction strength of both iso-scalar and iso-vector channels with an enlarged model space, we try to understand the uncertainties in the QRPA calculations. We also give reliable estimations for half-lives with newly calculated phase space factors from numerical electron wave functions \cite{SFW95}. These predicted half-lives are then compared to current experimental lower limits to explore the discovery potential of different nuclei. 

This article is arranged as follows, in Section II we give a brief introduction of the QRPA method and details of the NME and phase space factor calculations, in Section III we give detailed results followed by conclusions in section IV.

\section{Formalism}
The half-life of the $\beta\beta$-decay to $2^+_1$ states of daughter nuclei is expressed in a compact form as \cite{DKT85}:
\begin{eqnarray}
[\tau_{1/2}^{2\nu}(0^+\rightarrow 2^+)]^{-1}=G_{2\nu}^{2^+} g_A^4 |M^{2+}_{2\nu}|^2
\end{eqnarray}  
Where $G_{2\nu}^{2^+}$ is the phase space factor for the emitted electrons and neutrinos and $M^{2^+}_{2\nu}$ is the nuclear matrix element. Unlike normal conventions, we take the axial coupling constant $g_A$ outside the phase space factor.

The phase space factor (PSF) can be calculated by integrating over the lepton momenta \cite{DKT85,KI12}:
\begin{eqnarray}
G_{2\nu}^{2^+}=\frac{2\tilde{A}^6}{\ln 2}\int_{m_e}^{Q_{\beta\beta}^{2^+}+m_e} 
\int_{m_e}^{Q_{\beta\beta}^{2^+}+m_e-\epsilon_1}
\int_{0}^{Q_{\beta\beta}^{2^+}-\epsilon_1-\epsilon_2} \nonumber \\
\times f_{11}^{(0)} (\langle K_N \rangle-\langle L_N \rangle )^2 \omega_{2\nu} d \omega_1 d\epsilon_2 d\epsilon_1
\label{PhsSp}
\end{eqnarray}
where the energy denominator has the form:
\begin{eqnarray}
\langle K_N \rangle = \frac{1}{\epsilon_1+\omega_1 +\langle E_N \rangle -E_I} 
+ \frac{1}{\epsilon_2 + \omega_2 +\langle E_N \rangle -E_I}  \nonumber \\
\langle L_N \rangle = \frac{1}{\epsilon_1+\omega_2 +\langle E_N \rangle -E_I} 
+ \frac{1}{\epsilon_2 + \omega_1 +\langle E_N \rangle -E_I} \nonumber
\end{eqnarray}
Here $\epsilon_{1(2)}$ and $\omega_{1(2)}$ are energies of the outgoing electrons and neutrinos, respectively. The energy conservation requires that $\epsilon_1+\epsilon_2+\omega_1+\omega_2=Q+2m_e$ (we neglect the nuclear recoil energy). And $\langle E_N \rangle$ is a suitably chosen value for average excitation energies of the intermediate nucleus. The lepton kinematic factor $\omega_{2\nu}$ has the form:
\begin{eqnarray}
\omega_{2\nu}=\frac{(G\cos \theta_C)^4}{64\pi^4} \omega_1 \omega_2 p_1 p_2 \epsilon_1 \epsilon_2 \nonumber
\end{eqnarray}
And the closure energy $\tilde{A}$ is introduced to separate the nuclear and lepton parts \cite{DKT85}:
\begin{eqnarray}
\tilde{A}=\langle E_{N} \rangle +\frac{M_m-M_I}{2}+\frac{M_m-M_F}{2}= \langle E_{N}  \rangle + E_a\nonumber
\end{eqnarray}
An empirical formula for $\tilde{A}$ for double beta decay to the ground state can be found in \cite{KI12}. Here $E_a$ is the average mass differences between intermediate and initial (final) nuclei.

The function of electron radial wave functions (ERWF) $f_{11}^{(0)}$ is defined as:
\begin{eqnarray}
f^{(0)}_{11}=|f^{-1-1}|^2+|f_{11}|^2 + |f^{-1}{}_1|^2+|f_{1}{}^{-1}|^2
\end{eqnarray}
with
\begin{eqnarray}
f^{-1-1}&=&g_{-1}(\epsilon_1,R) g_{-1}(\epsilon_2,R) \nonumber \\
f_{11}&=&f_1(\epsilon_1,R) f_1(\epsilon_2,R) \nonumber \\
f^{-1}{}_{1}&=&g_{-1}(\epsilon_1,R) f_{1}(\epsilon_2,R) \nonumber \\
f_{1}{}^{-1}&=&f_1(\epsilon_1,R) g_{-1} (\epsilon_2,R) \nonumber
\end{eqnarray}
Here $g_{-1}$ and $f_1$ are upper and lower component of $s$-wave Dirac electron wave functions as defined in \cite{DKT85}. In this work, we follow the normalization in \cite{SFW95,KI12} for electron radial wave functions. We adopt the long wave length approximation (or so-called no finite de Broglie wave length correction in \cite{DKT85}) to separate the spatial and momenta integrations. It assumes that a constant lepton wave functions inside the nucleus with the constants chosen to be the values of $g_{-1}$ and $f_1$ at the nuclear surface ($r=R$, with {\it R} the nuclear radius $R=1.2A^{1/3}$fm) for electron. And to derive this phase space factor, we use the long wavelength approximation for neutrinos, that is only neutrino s-wave radial functions are nonzero ($j_0(kR)=1$).

The nuclear part of the decay, namely the nuclear matrix element (NME), depends on the details of the nuclear structure. It is known that the first $2^+$ states of the even-even nuclei are for spherical nuclei usually collective states of harmonic vibration. Quasi-particle Random Phase Approximation (QRPA), which well describes small amplitude harmonic vibrations of the spherical even-even nuclei, can be a reasonable approach for descriptions of such states. In this work, the QRPA method is used to construct both $1^+$ states of intermediate odd-odd nuclei (charge exchange version, named pn-QRPA) and $2^+$ excited states of final even-even nuclei (charge conserving (CC) version, namely CC-QRPA). QRPA starts with BCS or HFB vacua. Its basic ingredients, quasi-particles, are obtained by solving the BCS or HFB equations. The constructed excited states have the general forms $|J^\pi,m\rangle = Q_{J^\pi,m}^\dagger|0\rangle$ for intermediate nuclei and $| \mathcal{J}^\pi, m \rangle=\mathcal{Q}_{\mathcal{J}^\pi,m}^\dagger|0\rangle$ for the final nuclei. The creation operators $Q_{J^\pi,m}^\dagger$($\mathcal{Q}_{\mathcal{J}^\pi,m}^\dagger$) are superposition of two quasi-particle excitations, they are defined as\cite{BK95}:
\begin{eqnarray}
\mathcal{Q}_{\mathcal{J}^\pi,m}^\dagger &\equiv& \sum_{\tau \tau'}(\mathcal{X}_{m}^{\mathcal{J}^\pi} [\alpha_{\tau}^\dagger \alpha_{\tau'}^\dagger]_{\mathcal{J}^\pi}-\mathcal{Y}_{m}^{\mathcal{J}^\pi} [\tilde{\alpha}_{\tau} \tilde{\alpha}_{\tau'}]_{\mathcal{J}^\pi})  \nonumber \\
Q_{J^\pi,m}^\dagger&\equiv&\sum_{pn} (X_{m}^{J^\pi} [\alpha_{p}^\dagger \alpha_{n}^\dagger]_{J^\pi}-Y_{m}^{J^\pi} [\tilde{\alpha}_{p} \tilde{\alpha}_{n}]_{J^\pi})
\end{eqnarray}
Here $\tau$ and $\tau'$ have the same $\tau_z$ and can indicate to either protons or neutrons, for $\mathcal{Q}^\dagger$. $\alpha^\dagger$ is the quasi-particle creation operator connected with the single particle creation and annihilation operators by the BCS transformation $\alpha_i^\dagger=u_i c_i^\dagger+v_i \tilde{c}_i$, $\tilde{c}_i$ is the time reversed counterpart of the single particle annihilation operator $c_{i}$. The forward and backward amplitudes $X$($\mathcal{X}$) and $Y$($\mathcal{Y}$) can be obtained from solving the so-called QRPA equations (Here {\it X} and {\it Y} are amplitudes for the intermediate states and $\mathcal{X}$ and $\mathcal{Y}$ are amplitudes for the $2^+$ states):
\begin{eqnarray}
\left(
\begin{array}{cc}
A & B \\
-B^* & -A^*
\end{array}
\right)
\left(
\begin{array}{c}
X\\
Y
\end{array}
\right)=\omega \left(
\begin{array}{c}
X\\
Y
\end{array}
\right)
\end{eqnarray}
Here the interaction matrices $A$ and $B$ for CC-QRPA and pn-QRPA with realistic forces are expressed in Ref.\cite{SSF98}. One notices that for the particle-particle interactions, we have only the iso-vector ({\it T}=1) channel for CC-QRPA but   both iso-vector and iso-scalar channels for pn-QRPA. A strategy of parametrization of the renormalization strength in these channels will be discussed later.

With these calculated QRPA states, the nuclear matrix element (NME) for $2\nu\beta\beta$ to $2^+$ has the form \cite{DKT85}:
\begin{eqnarray}
M^{2+}_{2\nu}=\frac{1}{\sqrt{3}}\sum_{m_i,m_f}\frac{\langle 2_f^+||\sigma||1^+_{m_f}\rangle \langle 1^+_{m_f}|| 1^+_{m_i}\rangle \langle 1^+_{m_i} || \sigma || 0_i^+\rangle }{(E_a+(E_i+E_f)/2)^3} \nonumber \\
\end{eqnarray}
Here terms in the energy denominator are defined as $E_i=\omega_{m_i}-\omega^{i 1^+}_{1}+E_{1^+}^{exp.}$ and $E_f=\omega_{m_f}-\omega^{f 1^+}_{1}+E_{1^+}^{exp.}$, with $\omega^{i(f)1^+}$ being the lowest QRPA eigenvalues for $1^+$ intermediate states excited from initial and final nuclei. $E_{1^+}^{exp.}$ is the experimental excitation energy of the first $1^+$ state for the intermediate nucleus. 

The transition amplitudes from initial states to the intermediate ones are in our case:
\begin{eqnarray}
\langle m ||\sigma ||0^+_i\rangle
= \sum_{pn} \langle p ||\sigma || n\rangle (X^m_{pn} u_p v_n + Y^m_{pn} v_p u_n )
\label{init}
\end{eqnarray}

The overlap between the initial and final intermediate states can be written approximately as:
\begin{eqnarray}
\langle m_f || m_i \rangle&\approx& \sum_{pn} (X^{m_i}_{pn} X^{m_f}_{pn}-Y^{m_i}_{pn} Y^{m_f}_{pn})\nonumber \\
&\times&(u^{i}_p u^{f}_p+v^{i}_p v^{f}_p)(u^{i}_n u^{f}_n+v^{i}_n v^{f}_n)
\end{eqnarray}
Here we assume that the initial and final ground states are the same: $\langle BCS_i|BCS_f\rangle\approx 1$. For our BCS solution, the phase convention of positive u's and v's is used.

The transition strength from the intermediate to final $2^+$ states are more complicated \cite{CS94}
\begin{widetext}
\begin{eqnarray}
\langle 2^+_f ||\sigma ||m\rangle&=&\sqrt{15}\sum_{pn}\langle p || \sigma || n \rangle 
[\sum_{ p'\le p}\frac{(-1)^{j_{p'}+j_n}}{\sqrt{1+\delta_{p p'}}}  
\left\{
\begin{array}{ccc}
2 & j_{p'} & j_p \\
j_{n} & 1 & 1
\end{array}
\right\}
(u_p u_n \mathcal{X}_{p' p}^{2^+_f} X^{m}_{p'n}-v_p v_n \mathcal{Y}_{p' p}^{2^+_f} Y^{m}_{p'n})
\nonumber \\ 
&+& \sum_{p'\ge p}\frac{(-1)^{j_p+j_n}}{\sqrt{1+\delta_{p p'}}} 
\left\{
\begin{array}{ccc}
2 & j_{p'} & j_p \\
j_{n} & 1 & 1
\end{array}
\right\}
(u_p u_n \mathcal{X}_{p p'}^{2^+_f} X^{m}_{p'n}-v_p v_n \mathcal{Y}_{p p'}^{2^+_f} Y^{m}_{p'n})
\nonumber \\  
&-&\sum_{n'\le n}\frac{ (-1)^{j_n+j_p}}{\sqrt{1+\delta_{nn'}}} 
\left\{
\begin{array}{ccc}
2 & j_{n'} & j_n \\
j_{p} & 1 & 1
\end{array}
\right\} 
(v_p v_n \mathcal{X}_{n'n}^{2^+_f} X^{m}_{pn'} -u_p u_n \mathcal{Y}_{n' n}^{2^+_f} Y^{m}_{p n'} ) 
\nonumber \\
&-&
\sum_{n' \ge n}\frac{(-1)^{j_{n'}+j_p}}{\sqrt{1+\delta_{n n'}}} 
\left\{
\begin{array}{ccc}
2 & j_{n'} & j_n \\
j_{p} & 1 & 1
\end{array}
\right\}
(v_p v_n \mathcal{X}_{n n'}^{2^+_f} X^{m}_{p n'} -u_p u_n \mathcal{Y}_{n n'}^{2^+_f} Y^{m}_{p n'} ) ]
\nonumber \\
\end{eqnarray}
\end{widetext}
Here, $\mathcal{X}^{2^+}$ and $\mathcal{Y}^{2^+}$ are the forward and backward amplitudes of the first $2^+$ state of the final nucleus.

The expression of Fermi and Gamow-Teller NMEs for $2\nu\beta\beta$ to ground states can be found in literature \cite{SRF13} with a similar expression as eq.\eqref{init}. In this work, the NMEs for decays to ground states are used to determine the parameters of our method according to the different sensitivities of different parts of NME on different channels of the pp residual interaction \cite{RF11}: $g_{pp}^{T=0}$ is determined by experimental $2\nu\beta\beta$ GT NME and $g_{pp}^{T=1}$ by requiring a vanishing $2\nu\beta\beta$ Fermi NME.

\section{Results and discussions}
\begin{figure*}
\includegraphics[scale=0.6]{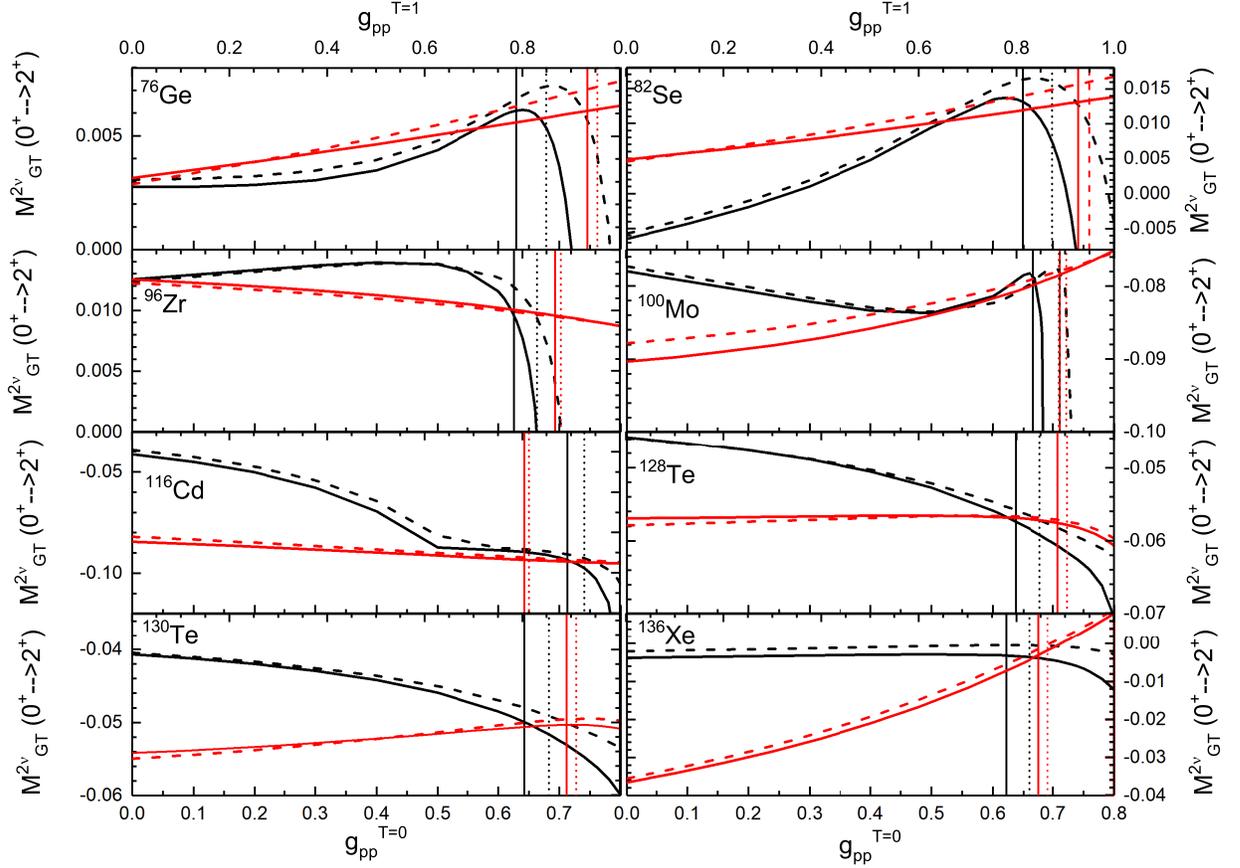}
\caption{(Color online) Dependence of NMEs for $2\nu\beta\beta$ to first $2^+$ excited states on the strength of iso-vector ($g_{pp}^{T=1}$) and iso-scalar ($g_{pp}^{T=0}$) pp residual interactions. Here the solid lines are for the larger model space and dashed lines for the smaller one as indicated in the text. The red curves are for iso-vector dependence and blacks for iso-scalar, the upper x-axes of each panels are for $g_{pp}^{T=1}$ and lower ones for $g_{pp}^{T=0}$, they are with different scale. And the vertical lines are for fitted $g_{pp}$'s with unquenched $g_A$ with line styles following the curves.}
\label{2npp}
\end{figure*}

\subsection{Nuclear Matrix Elements}

In this work we perform calculations for 8 nuclei which are supposed to be spherical and whose $2\nu\beta\beta$-decay half-lives are experimentally determined \cite{Bar15}: $^{76}$Ge, $^{82}$Se, $^{96}$Zr, $^{100}$Mo, $^{116}$Cd, $^{128}$Te, $^{130}$Te and $^{136}$Xe. The NMEs obtained from measured $2\nu\beta\beta$ half-lives are then used to determine the parameters in our calculations.

The general parametrization of this work can be summarized as follows. The single particle energies are obtained from solutions of Schr\"odinger equations with Coulomb corrected Woods-Saxon (WS) potentials and for the wave functions, we use the spherical Harmonic Oscillator (HO) wave functions (The advantage of using HO wave functions is that many s.p. matrix element can be analytically derived, a comparison between $\beta\beta$-decay calculations of using WS and HO has been done in \cite{FFR11}). For the model space, we adopt in this work two different sets for the sake of understanding the effect of model space truncations in our calculations. For the smaller one, we choose all single particle levels from {\it N}=0 up to one shell above the Fermi energy. For the larger one we add one more major shell. So for $^{76}$Ge and $^{82}$Se we have 21 s.p. levels ({\it N}=0-5) for the small model space (SMSp) and 28 levels ({\it N}=0-6) for the large one (LMSp), while for the other 6 nuclei, SMSp consists of 28 s.p. levels and LMSp 36 levels ({\it N}=0-7).

For the pairing part, we adopt the Br\"uckner G-matrix (of the CD-Bonn force), multiplied by the renormalized strength $g^{pair}$'s  to reproduce the experimental pairing gaps obtained from the five-point formula. As for QRPA, we use the same residueal interactions, with separate renormalized strength for both particle-hole ($g_{ph}$) and particle-particle ($g_{pp}$) channels respectively. For pn-QRPA, $g_{ph}$ is set to unity, while for CC-QRPA, $g_{ph}$ is fixed by reproducing the first $2^+$ excitation energies of final nuclei \cite{CS94}. We find that $g_{ph}$ in CC-QRPA deviates from unity, this is due to the anharmonicity beyond QRPA \cite{JZ12}. As has been shown \cite{RF11}, for the $2\nu\beta\beta$ to ground states, $M^{2\nu}_{GT}$ is sensitive to $g_{pp}^{T=0}$ only, while $M^{2\nu}_{F}$ is sensitive to the iso-vector part only. Therefore the parameters of $g_{pp}^{T=0}$ and $g_{pp}^{T=1}$ can be fitted by setting $M^{2\nu}_{GT}$ to be the experimental values and $M^{2\nu}_F$ to be zero respectively \cite{SRF13} as we indicated above. Thus for $2\nu\beta\beta$ to excited $2^+$ states, the only $g_{pp}$ parameter undetermined is the one for CC-QRPA in iso-vector channel. For consistency, it is natural to set it equal to that of pn-QRPA in iso-vector channel (a consistency check for iso-vector pp interactions in QRPA and pairing parts has been done in \cite{SRF13}). As a consequence, we now have only two $g_{pp}$ parameters in our calculation, $g_{pp}^{T=1}$ (for CC- and pn-QRPA) and $g_{pp}^{T=0}$(for pn-QRPA).

%As have shown in\cite{SRF13}, for $M^{2\nu}_F(0^+)$, the fermi component, its value decrease as the iso-vector strength $g_{pp}^{T=1}$ increases, meanwhile it remains constant when the iso-scalar strength $g_{pp}^{T=0}$ changes. For the GT component of $2\nu\beta\beta$, $M^{2\nu}_{GT}$, the dependence is of the opposite, it has an sensitive dependence on $g_{pp}^{T=0}$ and insensitive to $g_{pp}^{T=1}$. This was explained in\cite{RF11}. 

For $\beta\beta$-decay to $2^+_1$ states, only GT component is relevant, but due to the inclusion of final $2^+_1$ states described by $g_{pp}^{T=1}$ dependent CC-QRPA, the GT NMEs now depend on both iso-scalar and iso-vector pp interactions. Such dependence is illustrated in fig.\ref{2npp}, and it helps us to understand the uncertainties in our calculations. At the first glance of fig.\ref{2npp}, we find that for different nuclei, the difference of their NME values could be of more than a factor of 10, this differs drastically from decays to ground states \cite{Bar15} where the NMEs are basically within the same orders of magnitude. This is partly due to energy denominator's cubic dependence which heavily suppresses the NME with large intermediate energies, it also leads to the smallness of NMEs compared to the decay to ground states. On the other hand, the interplay between the pn-QRPA phonon and CC-QRPA phonon could also change the NME by orders of magnitude. 

The dependence of the NMEs for decay to excited states on iso-scalar pp channel (black curves in fig\ref{2npp}, where we keep $g_{pp}^{T=1}$ constant with the fitted value mentioned above) is much more complicated than that for the decay to ground states \cite{SRF13}. These curves suggest that when $g_{pp}^{T=0}$ approaches the values where QRPA equations collapse, the corresponding NME will drop rapidly to $-\infty$, this is similar to the decay to ground states (see {\it eg.} \cite{SRF13}). Besides, for $^{128}$Te, $^{130}$Te and $^{136}$Xe, we find similar trends for the $g_{pp}^{T=0}$ dependence between decays to excited and ground states, this may suggest something in common for the structure of their $2^+$ states and ground states. For $^{116}$Cd, similar decreasing trend of the curve is observed except a deceleration of such decrease after specific point.  The remaining nuclei show diversities for the $g_{pp}^{T=0}$ dependence. $^{76}$Ge and $^{82}$Se experience smoothing accelerating increase before the sudden decrease near QRPA collapse. Meanwhile, $^{96}$Zr sees a mild increasing before a rapid drop of NME. And $^{100}$Mo combines the behaviors of above nuclei. On the other hand, at $g_{pp}^{T=0}=0$, some NMEs are positive and other negative. There is actually a phase uncertainty of the NMEs since the measurements can determine only their absolute values and in this work we use the phase convention that forces the values of NME near QRPA collapse to be $-\infty$ to match the behavior of decays to ground states for the sake of comparison.

Opposite to the case for iso-scalar interactions, the NMEs change almost monotonically when $g_{pp}^{T=1}$, the iso-vector pp interaction strength changes (red curves). Effectively, the study of double beta decay concerns more about the absolute NME rather than actual NME, since the half-lives depend on the square of NME. With this respect, we find that some nuclei get reduced decay strength with increasing $g_{pp}^{T=1}$ but others such as $^{76}$Ge get enhanced strength. As for the magnitude of the changes by varying $g_{pp}^{T=1}$, it is usually smaller than that of $g_{pp}^{T=0}$ except $^{136}$Xe. For iso-vector interaction, sharp drops of NMEs are not observed. This is because the collapse of QRPA happens with $g_{pp}^{T=1}$ much larger than realistic values. 

%The small sensitivity of NME of $g_{pp}^{T=1}$ leads to the fact that our treatment of iso-spin symmetry restoration will change the absolute NMEs for most nuclei by just several percents. This is unlike the case for $0\nu\beta\beta$, where the fermi matrix elements are reduced by more than 10\%. 

Fig.\ref{2npp} also shows that the size of model space doesn't affect the evolving behavior of NME, although it does change their actual values for specific $g_{pp}$'s. As usual, for a smaller model space, larger fitted $g_{pp}$ values are expected. However, for most cases, the resulting NMEs don't change too much. This suggests that the model space truncation affects the NMEs of decays to ground states and excited states by the same way. Thus, the results with SMSp and LMSp differs only by several percents, but for nuclei whose NMEs at realistic $g_{pp}$ values are near zero ($^{136}$Xe), the truncation of model space may lead to large relative changes of the NMEs.% in the case the NMEs itself is small in magnitude.
%In general, to restore the isospin symmetry, we readjusted the values of $g_{pp}^{T=1}$ compared to treatments in\cite{BK95,SSF97}, this would lead to the change of NME's by about several percents to several tens of percents depending on nucleus.

\begin{figure*}
\includegraphics[scale=0.32]{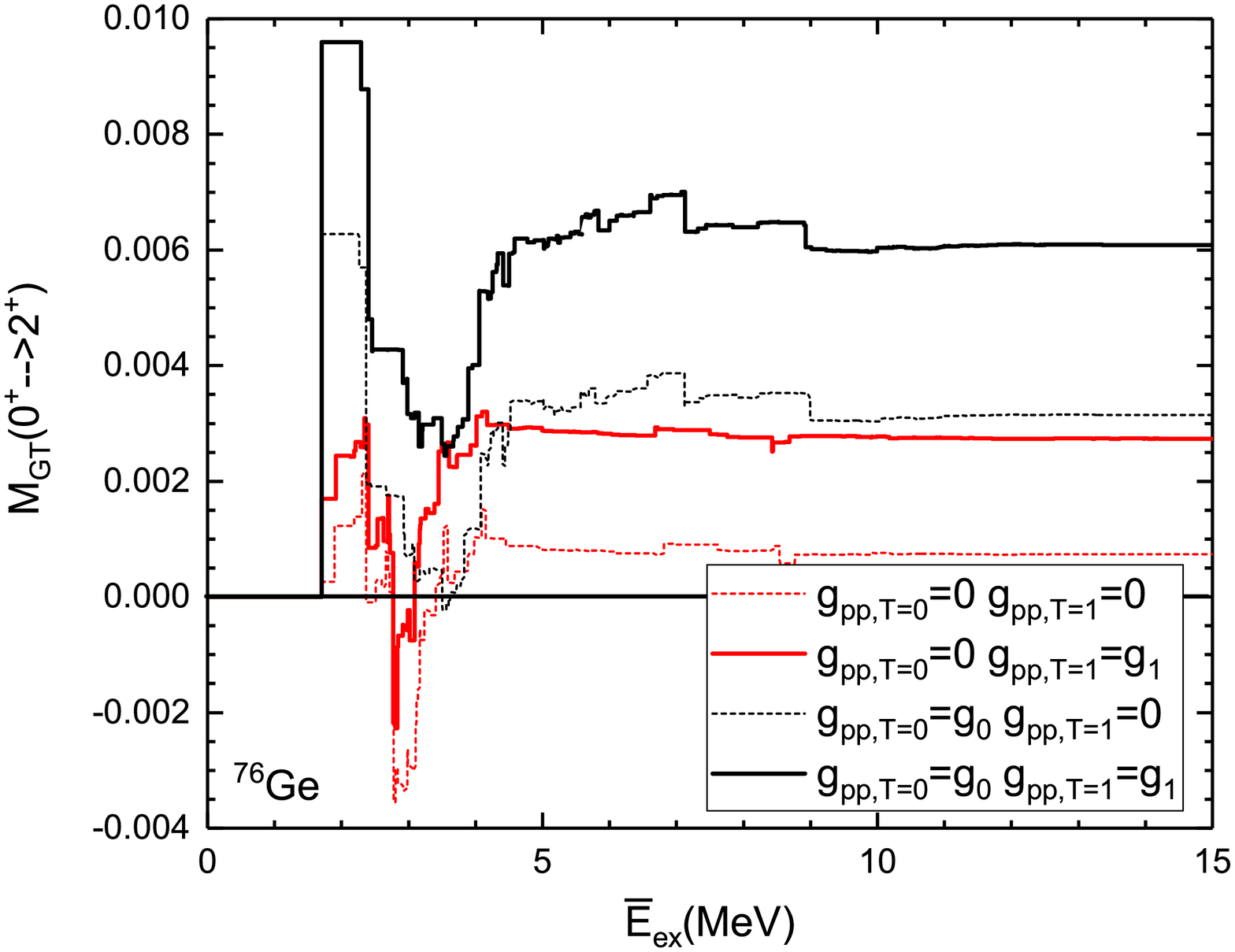}
\includegraphics[scale=0.32]{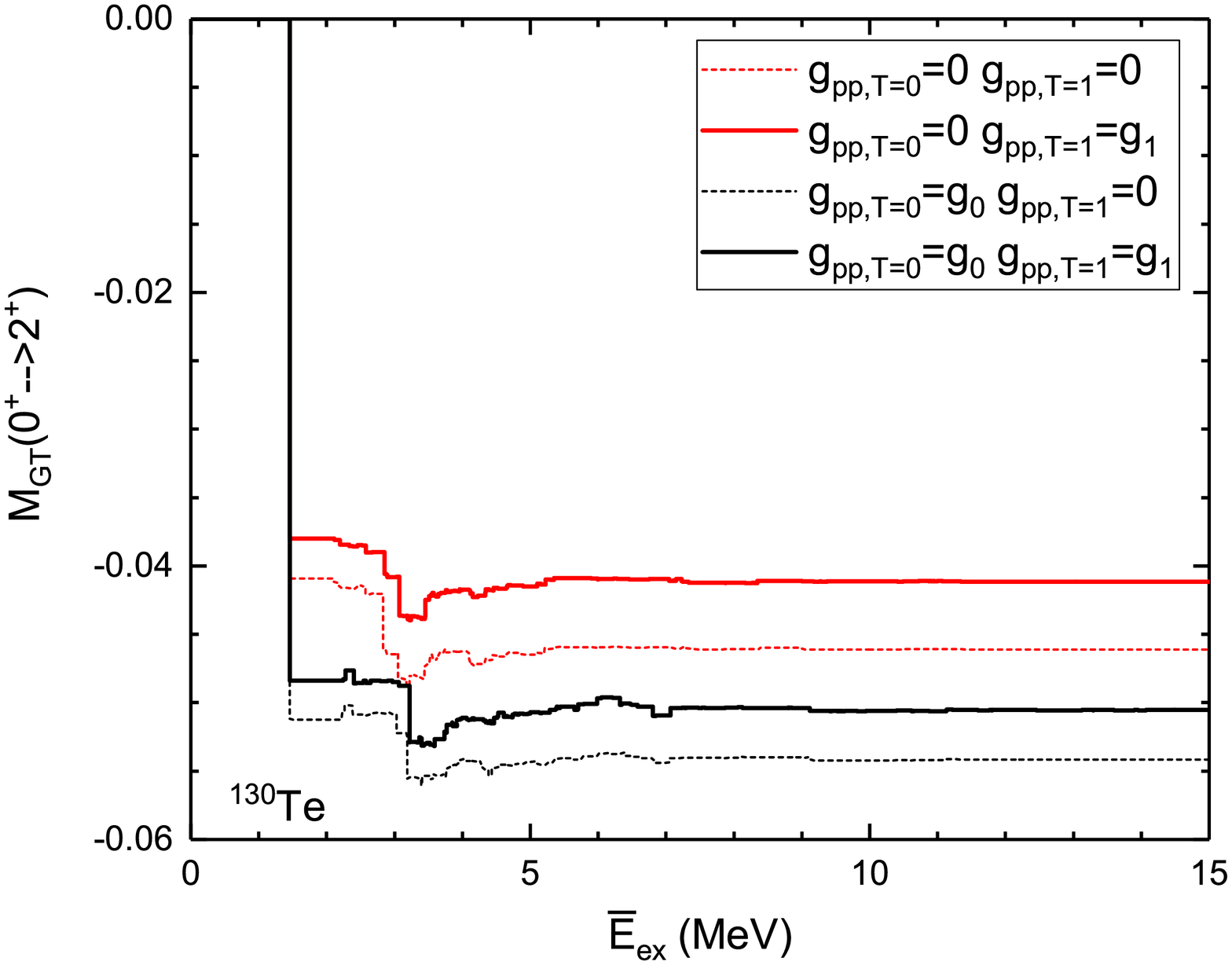}
\caption{(Color online) Running sum of $2\nu\beta\beta$ to $2^+_1$ for $^{76}$Ge and $^{130}$Te. Here $g_{0}$ and $g_{1}$ are fitted $g_{pp}$ values for T=0 channel and T=1 channel respectively with unquneched $g_A$.}
\label{runsum}
\end{figure*}

Our results can be compared to results obtained from other QRPA calculations \cite{AS96,BK95}, where $g_{pp}^{T=1}$ and $g_{pp}^{T=0}$ haven't been separated and smaller model spaces are used. Since the evolving trend of $g_{pp}^{T=1}$ is monotonic, the combined evolving trends of $g_{pp}$ are more close to that of $g_{pp}^{T=0}$. We find that we have similar trends as other calculations but we have differed NME values from them. Meanwhile different calculations deviate with each other, a large discrepancy exists for different calculations. If we consider calculations from other approaches, discrepancies could be further magnified. These discrepancies can only be solved by measurements. From fig.\ref{2npp}, we learn that the treatment of isospin symmetry restoration ($g_{pp}^{T=0}\ne g_{pp}^{T=1}$) could alter the final NMEs within several percents by most nuclei except for $^{136}$Xe. Since the final NMEs for $^{136}$Xe are crossing zero in the relavant $g_{pp}$ region, old isospin violating treatment will give a fairly large NME.

To understand the behavior of NME's dependence on $g_{pp}$, one could look into the running sum which shows how the low- and high-lying states contribute. We plot in fig.\ref{runsum} two typical cases, $^{76}$Ge and $^{130}$Te which we discussed above. At first, the running sum shows more complicated structure than $\beta\beta$-decay to ground states \cite{CS99,FFR09}. In \cite{FFR09}, low-lying states contribute mostly positively and contributions from high-lying states may give enhancements or reductions to total NMEs according to the values of $g_{pp}$. However, in calculations of the decay to $2^+_1$, it is quite different, the high-lying contributions are largely suppressed, the effect starts to appear only near the collapse of QRPA. At realistic $g_{pp}$ region, most of the contributions are from low-lying transitions and they may add up together or cancel each other from case to case.

Iso-vector pp interactions doesn't change the basic structure of the running sums, but it slightly changes the magnitude of each transitions, this can be seen by comparing the solid and dashed curves. For each nuclei, we could always find corresponding transitions between cases with and without iso-vector pp interaction. These transitions are with similar energies but differ by the absolute magnitude of transition strength. For both low- and high-lying strength, we can also find increasing strength with the iso-vector strength. %And the slope of the NME-$g_{pp}^{T=1}$ curve depends on the competitions over which transitions grow faster. % to be checked The difference of NME dependence on $g_{pp}^{T=1}$ are comes from 

The effect of pp interactions in the iso-scalar channel to the NMEs can be acquired by comparing red and black curves. When $g_{pp}^{T=0}$ is zero, barely no high-lying states contributions are observed, with large $g_{pp}^{T=0}$, we see moderate reductions from high-lying states near 10MeV (The proposed GTR region) for $^{76}$Ge. This reduction from high-lying states becomes drastic near QRPA collapse in our calculation (For $^{130}$Te, this effect is not shown in the sum rule because the fitted $g_{pp}^{T=0}$ is far away from the collapse, and near collapse, we could observe the reduction around GTR energies). This leads to the drastic decrease of NMEs at very large $g_{pp}^{T=0}$ in fig.\ref{2npp}. From fig.\ref{runsum}, we could also understand different behavior of NME's on the $g_{pp}^{T=0}$ dependence of various nuclei. For example, for $^{76}$Ge, low-lying strength strongly cancel each other, while the positive strength grow faster than negative ones with increasing $g_{pp}^{T=0}$, this causes the accelerated increase until the reduction from high-lying states mentioned above dominates the NME. This contradicts the $^{130}$Te case where major strength is all negative and adds up together, so no increase of the strength is observed; the contribution from high-lying states (not shown in the graph) further reduces the results.  At the low-lying energy region, iso-scalar pp interactions changes not only the strength of each state but also the structure of the running sum, this produces the complicated evolution of NME. %As for the case of $^{116}$Cd, the excitation energy of the leading contribution is decreasing until it becomes the first $1^+$ energy, due to the strategy of our choice of the energy denominator. This means the energy denominator of this very strong transition from a decrease as function of $g_{pp}^{T=0}$ to an constant at transition point. This leads to non-smooth change of the curves.

%For the sake of comparison, we present both the small and large model space as mentioned above. We then come to the conclusion that, the addition of one more major shell in our calculations doesn't change too much the final NME results, although the fitted $g_{pp}$'s changes apparently.  

\subsection{Phase Space Factors and Half-Lives}

\begin{figure}
\includegraphics[scale=0.32]{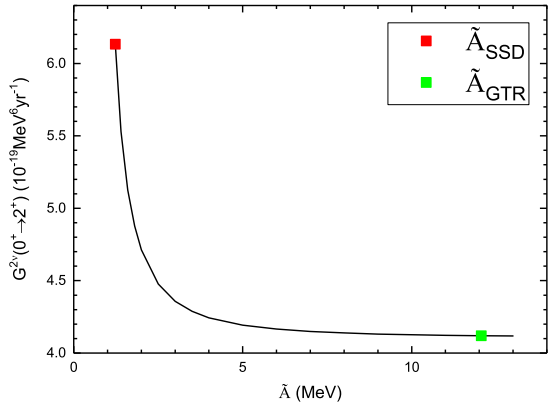}
\caption{(Color online) Phase space factor for $2\nu\beta\beta$ to $2^+_1$ of $^{116}$Cd as a function of average excitation energy $\tilde{A}$, where $\tilde{A}_{SSD}$ corresponds to the Single state dominance case where the average excitation energy of intermediate nucleus is $0$, and for $\tilde{A}_{GTR}$ the average excitation energy of intermediate nucleus is taken to be the GTR energy from \cite{KI12}.}
\label{psfe}
\end{figure}

PSFs of the decay to the $2^+_1$ state have been addressed in several publications \cite{DKT85,PNM14,Sto19}. In this article, we calculate the PSF with numerical electron wave functions from the numerical package {\it RADIAL} \cite{SFW95}, to avoid complication, we assume the electron wave functions is constant inside the nucleus. This yields a separation of the calculations of NMEs and the PSFs as discussed above. We use uniformly distributed electric charge in the nuclei to take into consideration the nuclear finite size \cite{KI12}, and the charge radius is taken to be the empirical nuclear radius $R=1.2A^{1/3}$fm. We neglect the screening effect from atomic electrons. Their effects are analyzed in references such as \cite{Sto19}. To separate the nuclear and lepton parts, one uses the average excitation energies $\tilde{A}$ in the phase space calculations (see eq.\eqref{PhsSp}). There is always arbitrariness of the choice for $\tilde{A}$ in such a formalism. To estimate a possible error due to this choice, we plot the dependence of this phase space factor on $\tilde{A}$ for $^{116}$Cd in fig.\ref{psfe}, the curve starts at lowest possible average energy ($\tilde{A}_{SSD}$) which is experimentally known. Here SSD is the abbreviation of "single state dominance", and this single state usually refers to the first $1^+$ state of intermediate states. As one can see, the PSF gets its largest value at the point of $\tilde{A}_{SSD}$ and then it rapidly drops to a nearly constant value at the average energy around 5MeV, after that it barely changes. $^{116}$Cd decays to excited state of $^{116}$Sn. The value changes about $30\%$, this is close to the case of the decay to ground states \cite{KI12}. 

\begin{table*}[htp]
\begin{center}
\label{PSF}
\caption{The calculated phase space factors from \cite{DKT85} and this work, decay half-lives for $\beta\beta$-decay to the first $2^+$ states from this work, where the queching factor $g_A=0.75 g_{A0}$ is used. The experimental half-lives or half-life limits of $2\nu\beta\beta$ to ground states and first $2^+$ states are tabulated as well. Here $Br(2^+)$ is the branch ratio of the decay to $2^+$ to overall $2\nu\beta\beta$. }
\begin{tabular}{|c|c|c|cccccc|c|}
 \hline
 		&$t^{2\nu,exp}_{1/2}(0^+_1)$(yr)  &Q&\multicolumn{3}{c}{ $G^{2\nu}_{2^+_1}$ (yr$^{-1}\cdot$ MeV$^6$)}  & $M^{2\nu}_{2^+_1}$ & $t^{2\nu,theo}_{1/2}(2^+_1)$ & $t^{2\nu,exp}_{1/2}(2^+_1)$  & $Br(2^+_1)$ \\
		 &  \cite{Bar15}&MeV& \cite{DKT85}   & HSD & SSD &  MeV$^{-3}$ & yr		&yr \cite{Bar17}&		\\
 \hline
$^{76}$Ge& $1.65^{+0.14}_{-0.12}\times 10^{21}$& 1.480	& 8.620$\times 10^{-24}$ 
& 7.599$\times10^{-24}$	& 1.053$\times10^{-23}$	& 6.08$\times 10^{-3}$
&3.33$\times 10^{27}$ 	&$>1.6\times 10^{23}$ 	&5.0$\times 10^{-7}$\\
 \hline
$^{82}$Se&$(9.2 \pm 0.7)\times 10^{19}$& 2.219	&1.569$\times 10^{-21}$
& 1.354$\times 10^{-21}$ 	& 3.408$\times 10^{-21}$	& 1.31$\times 10^{-2}$
&1.96$\times 10^{24}$ 	& $>1.0\times 10^{22}$ 	& 4.7$\times 10^{-5}$\\
  \hline
$^{96}$Zr&$(2.3 \pm 0.2)\times 10^{19}$& 2.572	&		-			
&1.407$\times 10^{-20}$	&1.935$\times 10^{-20}$	& 1.16$\times 10^{-2}$
&4.67$\times 10^{23}$   	& $>7.9\times 10^{19}$  	& 4.9$\times 10^{-5}$\\
\hline
$^{100}$Mo&$(7.1 \pm 0.4)\times 10^{18}$& 2.495	&1.382$\times 10^{-20}$	
&1.127$\times 10^{-20}$	&2.989$\times 10^{-20}$	& -7.83$\times 10^{-2}$  
&6.63$\times 10^{21}$	&$>2.5\times 10^{21}$	& 1.1$\times 10^{-3}$\\
\hline
$^{116}$Cd&$(2.87 \pm 0.13)\times 10^{19}$&	1.520 &	-			
&4.120$\times 10^{-23}$ 	&6.156$\times 10^{-23}$ 	& -9.05$\times 10^{-2}$  
&2.41$\times 10^{24}$	& $>2.3\times 10^{21}$	& 1.2$\times 10^{-5}$\\
\hline
$^{128}$Te&$(2.0 \pm 0.3)\times 10^{24}$& 0.423	&2.350$\times 10^{-29}$
&1.779$\times 10^{-29}$	&1.813$\times 10^{-29}$	& -5.72$\times 10^{-2}$  	
&2.05$\times 10^{31}$	& - 				  	& 9.8$\times 10^{-8}$\\
\hline
$^{130}$Te&$(6.9 \pm 1.3)\times 10^{20}$& 1.991	&2.119$\times 10^{-21}$
& 1.581$\times 10^{-21}$	& 2.713$\times 10^{-21}$	& -5.00$\times 10^{-2}$  
&1.79$\times 10^{23}$	& $>2.8\times 10^{21}$	& 3.9$\times 10^{-3}$\\
\hline
$^{136}$Xe&$(2.19 \pm 0.06)\times 10^{21}$& 1.639 &2.659$\times 10^{-22}$
& 1.755$\times 10^{-22}$	& 5.179$\times 10^{-22}$	&-3.04$\times 10^{-3}$  
&2.54$\times 10^{26}$	&$>4.6\times 10^{23}$	&8.6$\times 10^{-6} $\\
\hline
\end{tabular}
\end{center}
\end{table*}

For PSF calculations, there are usually two kinds of choices for $\tilde{A}$, SSD mentioned above and high-lying state dominance (HSD). Here the high-lying state usually refers to the strong Gamow-Teller Resonance (GTR) which is observed in charge-exchange experiments and its position can be obtained from systematics \cite{KI12}. For optimal choices of $\tilde{A}$, one could resort to nuclear structure data. Our analysis above about the running sum suggests that for decays to $2^+$, low energy states especially the first states make the largest contribution to the NME, therefore we choose $\tilde{A}_{SSD}$ in this work. These calculated PSFs are tabulated in Table \ref{PSF}. We have also tabulated the previous results from Ref.\cite{DKT85} which uses HSD ($\tilde{A}=10MeV$ for all nuclei) for estimation of PSF, we present also our calculated PSFs from HSD for the sake of comparison. By comparing our results with Ref.\cite{DKT85}, we find that two sets of results are generally within the same order of magnitude. The deviations between the current calculations and previous one are within a factor of 2, our current calculations with numerical electron wave functions yield smaller PSFs, the reductions vary from $10\sim40\%$. The reason for such an overestimation from previous work is that they use the constant electron wave functions with the values at the center of nuclei and we use the values at the surface. Such a choice  at the nuclear surface also comes from implications of nuclear structure calculations, such as those for single-$\beta$ decay in \cite{Fan19} and for $\beta\beta$-decay in \cite{SDS15}. As shown in fig.\ref{psfe}, the SSD PSFs corresponds to larger PSFs and therefore short half-lives. The errors from the choice of $\tilde{A}$ for the PSF depends on the Q values as well as the $1^+$ intermediate states energies. The reduction for HSD to SSD can be as large as $60\%$ for certain nuclei, but it can also be small, such as for $^{128}$Te with a much smaller PSF.

The half-life of the decay to the first $2^+$ can be obtained by combining the calculated NME and PSF.  Here for the 8 nuclei involved, the largest NME comes from $^{116}$Cd and the smallest comes from $^{136}$Xe, the difference is more than a factor of 10. The difference among PSFs is much larger. Three orders of magnitude deviations are observed by most nuclei mostly due to their different Q values and an extremely small PSF is found for $^{128}$Te who has a small Q value. 

The half-lives of these nuclei cross a large region, from $\sim 10^{22}$ to $\sim 10^{32}$ years. $^{100}$Mo has the shortest half-life for both $2\nu\beta\beta$-decay to ground and $2^{+}_1$ states, also it has a fair large decay branching ratio. This suggests its $\beta\beta$-decay to the $2^+_1$ state has the largest potential to be detected. While $^{128}$Te with a half-life of $10^{32}$ years seems impossible for detections. The same happens for $^{76}$Ge and $^{136}$Xe too. These three nuclei have extremely low branching ratios of the decay to the $2^+$ to the overall $\beta\beta$-decays either due to small PSF ($^{128}$Te) or due to small NME ($^{76}$Ge and $^{136}$Xe). $^{96}$Zr, $^{130}$Te and $^{100}$Mo are promising candidates for future experiments. Especially for $^{100}$Mo, the current experimental limit is close to the predicted half-life with less than one order of magnitude difference. With future improvements of experiments, it is perhaps possible to observe the special $2^+_1$ mode of this nucleus, since it is also the first nucleus for which the decay to the first $0^+$ excited was observed. For other nuclei, observations of $\beta\beta$-decay to $2^+_1$ can be quite difficult due to their long half-lives and small branching ratios.

\section{Conclusion and Outlook}
In this work, we calculated the nuclear matrix elements and phase space factors of $2\nu\beta\beta$ to the first $2^+$ states for 8 nuclei with partially restored isospin symmetry. We studied the NME dependence on the iso-vector and iso-scalar pp residual interaction strength. And finally we give predictions of half-lives and branching ratios of these decays to excited states. However, further investigation on the effect of anharmonicity of the $2^+$ phonon to the decay is needed.

\section*{acknowledgement}
This work is supported by National Natural Science Foundation of China under Grant No. 11505078 and 1164730,  "Light of West China" Program and key research program (XDPB09-2) from Chinese Academy of Sciences. We would like to thank Prof. F. \v{S}imkovic for useful discussions and help on the code.

\end{document}